\documentclass{elsart}
\usepackage{graphicx}

\def\bfg{\begin{figure}}
\def\efg{\end{figure}}
\begin{document}
\begin{frontmatter}
\title{Price Drops, Fluctuations, and Correlation in a Multi-Agent
Model of Stock Markets}

\author{A. G. Zawadowski$^1$, R. Kar\'adi$^1$, and J. Kert\'esz$^{1,2}$}

\address{$^1$Department of Theoretical Physics, Budapest University of Technology and Economics,
Budafoki \'ut 8, H-1111, Budapest, Hungary}
\address{$^2$Laboratory of Computational Engineering, Helsinki
University of Technology, P.O.Box 9400, FIN-02015 HUT, Finland}

\begin{abstract}
In this paper we compare market price fluctuations with the
response to fundamental price drops within the Lux-Marchesi model
which is able to reproduce the most important stylized facts of
real market data. Major differences can be observed between the
decay of spontaneous fluctuations and of changes due to external
perturbations reflecting the absence of detailed balance, i.e., of
the validity of the fluctuation-dissipation theorem. We found that
fundamental
price drops are followed by an overshoot with a rather robust
characteristic time.

\end{abstract}

\end{frontmatter}
\section{Introduction}

In the recent years physicists have shown increasing interest in
examining the statistical properties of real market financial data
\cite{paul} and they have contributed to the extraction of the most important
characteristics which are referred to as "stylized facts"
\cite{facts}. Such stylized facts include fat tailed distribution
and short time correlations for the logarithmic returns, volatility clustering,
gain-loss asymmetry, etc.

To deepen our understanding of financial markets building models
is essential. One approach is constructing purely mathematical
models (e.g., ARCH and GARCH processes which are well known to and
widely used by economists \cite{farmer}). Another way which is
more appealing for statistical physicists is that of the so-called
multi-agent models. These are based on interacting agents using
different strategies corresponding to real market behavior. The
simplest of these models are probably the so-called "minority
games" (e.g., the "El Farol bar" model \cite{elfarol}). Such
rather abstract models are appropriate to capture main mechanisms
but a detailed correspondence to economics is hard to find.  An
example for more complicated models based on both economical and
physical approaches is that introduced by Lux and Marchesi
\cite{lux1,lux2}. In this model a relatively large number of
parameters enables to incorporate several aspects of real
financial processes.

It is well known that prices on financial markets (particularly on
stock markets) tend to fluctuate in a broad range. It has for long
been a major goal of economists to understand the cause of these
fluctuations, rises, and drops \cite{cutler}. Experts seem to be
puzzled by the fact that sometimes price changes can be easily traced
back to well defined external effects like news about political
events, announcements of dramatic economic data etc., while in many
cases there is no apparent reason for the major fluctuations.

In statistical physics we distinguish between statistical
fluctuations and changes due to external perturbations. If time
reversal symmetry (or detailed balance) holds for the system like
at thermal equilibrium, the famous fluctuation-dissipation theorem
implies that spontaneous fluctuations and the response to small
perturbations decay in the same way. Of course, the condition of
detailed balance does not hold for financial markets. Even
statitionarity can be questioned and the time reversal symmetry is
broken, e.g., agents want to maximize profit and tend to switch to
winning strategies.

The goal of our study is to compare fluctuations and response to
external effects on financial markets. In order to have a clear cut
situation we investigate the the Lux-Marchesi model.
 The ability of the model to reproduce important
stylized facts together with the relative robustness of this property
when changing the parameters induced us to compare the fluctuations
and response (to fundamental price changes) within this framework. We
sincerely hope that our major results hold for real market data as
well.

\section{Simulation of the Lux-Marchesi model}

Let us start with summarizing the basic features of the
Lux-Marchesi single asset -- multi agent model. One of the main
assumptions is that there exists a fundamental price $p_f$ of
stocks (the value of the company and its prospective future
growth) around which the real price fluctuates. In this model the
agents are let to choose among the three following strategies:
optimists (who buy whatever happens), pessimists (who sell), and
fundamentalist (who sell if the market price is above the
fundamental price and vice versa). Optimists and pessimists
together are called chartists according to the usual terminology.

The number of all agents is $N=500$, of which the number of
fundamentalists is $n_f$, that of optimists $n_+$, $n_-$ for
pessimists, and $n_c=n_++n_-$ for chartists. The opinion index:
$x={{n_+-n_-}\over{n_c}}$ measures to what extent optimistic
strategy dominates among the chartists. The aggregate excess 
demand of the agents for the stocks is computed as
following:

$$ED = (n_+ - n_-)*t_c + n_f*\gamma*(p_f-p)$$

\vskip1cm
\begin{center}
\begin{tabular}{|c|c|}
\hline {\bf parameter} & {\bf description} \\ \hline $N$ & number
of all agents \\ $\nu_1$ & frequency of revaluation of opinion \\
$\nu_2$ & frequency of transition between fund. and chart. \\
$\beta$ & weight of demand   \\ $T_c=N*t_c$ & trade volume of
chartists \\ $T_f=N*\gamma$ & trade volume of fundamentalists
\\ $\alpha_1$ & strength of herding effect
\\ $\alpha_2$ & importance of price trend for chartists  \\ $\alpha_3$ &
importance of profit difference in transition\\ $s$ & discount factor of
fundamentalist profit
\\ $\sigma$ & standard deviation of $\mu$
\\ $R$ & average real return of other investments
\\ $r$ & nominal divident of the asset
\\ $dt$ & time increment in one simulation step
\\ $dt'$ & time increment (fast price change)
\\ $\Delta t$ & length of followed price trend
\\ \hline
\end{tabular}
\end{center}
\vskip0.5cm
\centerline{\protect\footnotesize Tab. 1: The description of the
parameters of the Lux-Marchesi model \cite{lux2}}

If the excess demand plus a small Gaussian noise term $\mu$ is
positive the market-maker may increase the price by $0.01$ with
the probability $\pi_{\uparrow p}$; if it is negative, the price
is adjusted downwards with the probability $\pi_{\downarrow p}$,
where these probabilities are:

$$\pi_{\uparrow p}= max [0, \beta (ED + \mu )] $$

$$\pi_{\downarrow p}= -min [0, \beta (ED + \mu )] $$

The dynamics of the model is governed by the rule that agents may
switch between the  strategies if prospective payoffs are better
using another strategy: the bigger the difference between the
payoffs the higher the probability that the agent switches to the
better strategy (transition probabilities are an exponential
function of the profit difference). The transition probabilities
between any two groups of traders are the following (where +
stands for optimists, - for pessimists, and f for fundamentalists:
thus e.g. $\pi_{+-}$ is the transition probability of an optimist
to pessimist during one time unit):

optimist -- pessimist:

$$\pi_{+-}=\nu _1\left( \frac {n_c}{N}\; exp(U_1)\right) $$
$$\pi_{-+}=\nu _1\left( \frac {n_c}{N}\; exp(-U_1)\right)  $$

where $$U_1=\alpha _1 x + \alpha _2 \frac {\dot p}{\nu _1}$$

optimist -- fundamentalist:

$$\pi_{+f}=\nu _2\left( \frac {n_+}{N}\; exp(U_{2,1})\right) $$

$$\pi_{f+}=\nu _2\left( \frac {n_f}{N}\; exp(-U_{2,1})\right)  $$

where $$U_{2,1}=\alpha _3 \bigg( \left( r + \frac {\dot p}{\nu
_2}\right) / p - R - s \bigg|  \frac {p_f - p}{p}\bigg|
\bigg)$$

pessimist -- fundamentalist:

$$ \pi_{-f}=\nu _2\left( \frac {n_-}{N}\; exp(U_{2,2})\right) $$
$$\pi_{f-}=\nu _2\left( \frac {n_f}{N}\; exp(-U_{2,2})\right) $$

where $$U_{2,2}=\alpha _3 \bigg( R - \left(r + \frac {\dot p}{\nu
_2} \right) / p - s \bigg| \frac {p_f - p}{p}\bigg| \bigg)
$$

The model uses many parameters, some of them of economic origin,
some of them determining the size and reaction speed of the market
(Tab. 1). The past price trend  is $\dot p(t)={p(t)-p(t+\Delta t )
\over \Delta t}$. In the simulation the  price and the number of
agents in each group of traders  is updated after a small time
interval $dt$, if the price change was rapid in the past couple of
simulation steps the elementary time step is reduced to $dt'$.

In our simulation we used the parameter sets given in the original
article of Lux and Marchesi (Tab. 2), if no other indication is
given parameter set IV was used for the simulation.

\vskip1cm
\begin{center}
\begin{tabular}{|c|c|c|c|c|}
\hline {\bf par. set:} & {\bf I} & {\bf II} & {\bf III} & {\bf IV}
\\ \hline $N$ &500& 500 &500 &500 \\  $\nu_1$& 3& 4& 0.5& 2  \\
$\nu_2$ &2& 1& 0.5 &0.6
\\  $\beta$& 6& 4& 2 &4 \\  $T_c$ &10 &7.5 &10
&5  \\  $T_f$ &5 &5 &10 &5  \\  $\alpha_1$ &0.6& 0.9& 0.75& 0.8 \\
$\alpha_2$& 0.2& 0.25& 0.25& 0.2 \\ $\alpha_3$& 0.5& 1 &0.75& 1
\\  $p_f$& 10& 10 &10& 10 \\  $r$ &0.004& 0.004& 0.004 &0.004 \\
$R$& 0.0004& 0.0004& 0.0004& 0.0004
\\  $s$& 0.75& 0.75& 0.8& 0.75 \\  $\sigma$& 0.05&
0.1& 0.1& 0.05 \\ $dt$ & 0.01& 0.01& 0.01&0.01 \\ $dt'$ & 0.002&
0.002& 0.002& 0.002
\\ \hline
\end{tabular}
\end{center}
\vskip0.5cm
\centerline{\protect\footnotesize Tab. 2: Values of the parameters
in the four different parameter sets \cite{lux2}}

\begin{figure}[ht] \centerline{
\includegraphics[width=14.0truecm]{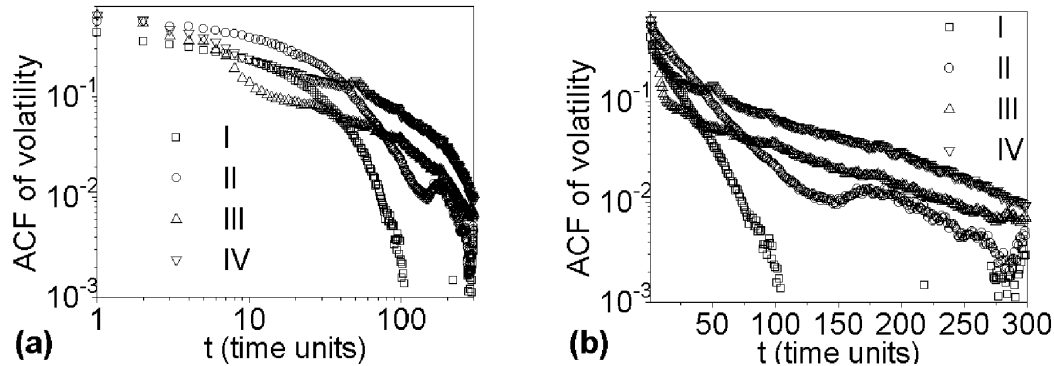}}
\caption{{\protect\footnotesize Autocorrelation function of
volatility for four different parameter sets (simulation length: 2
million time units) on (a) log-log and (b) lin-log scale.}}
\label{acf}
\end{figure}

During our work we found that in the Lux-Marchesi model the
autocorrelation of volatility shows exponential decay instead of a
power-law time dependence in contradiction to the general view on
real market data \cite{paul,liu}.  The power-law decay is usually
attributed to some kind of scale-invariance (regarding time) in
financial markets (Fig. \ref{acf}) implying that real markets are
in a critical state, which we could not find in the Lux-Marchesi
model even though we tried varying the parameters in a broad
range. Nevertheless, the phenomenon of volatility clustering is
described by the model resulting in large characteristic times of
the autocorrelation function which show that the model captures
important aspects of the market. Some other problems regarding the
thermodynamic limit (TDL) of the model were pointed out earlier
\cite{egenter}. We think that these are irrelevant from our point
of view since real markets (and our simulations) are far from the
TDL and the time scale of the mentioned effect is bigger then that
of studied fluctuations and response.

\section{Spontaneous price fluctuations}

The dynamics of the model implies that there are continuous
fluctuations around the fundamental price. We examined the decay
of these fluctuations using the following simple method: when the
price rose to $p=p_f+\Delta p=10.0+\Delta p$ we defined this as a
fluctuation and observed the average decay for many runs. Hence we
did not try to determine whether the price really sank after
reaching $p=p_f+\Delta p$ assuming that (at least at bigger
fluctuations) the probability of further rise is much smaller than
that of further decline.

\begin{figure}[ht]
\centerline{
\includegraphics[width=14.0truecm]{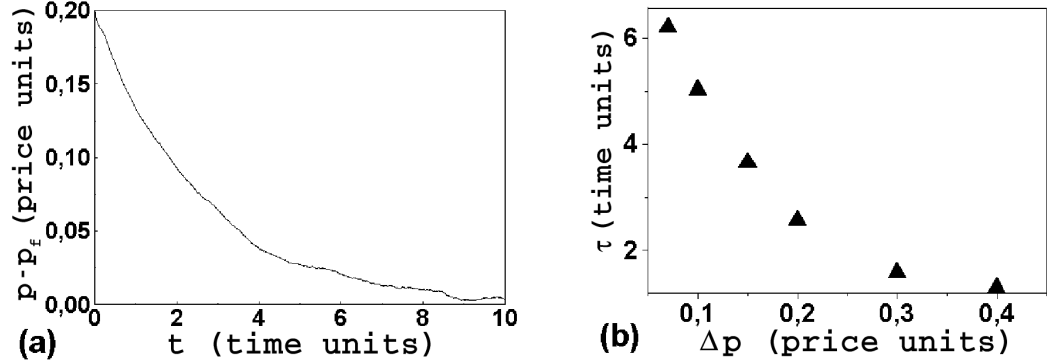}}
\caption{{\protect\footnotesize (a) Average decay of 8425 price
drops from $10.2$ to $10.0$ price units as a function of time, (b)
average characteristic time of decay as a function of the size of
decay (data points computed as an average of 7625 to 120529
fluctuations) }} \label{flukt1}
\end{figure}

We observed exponential decay for the price fluctuations using all
parameter sets in accordance with the well known fast decay of the
correlation function. Detailed simulations were undertaken for
parameter set IV (Fig. \ref{flukt1}a). Furthermore, in case of
relatively large fluctuations we observed that the opinion index
and the fraction of chartists significantly differed from their
average values and exponential decay was observed for both
quantities. The characteristic time of the decay of the price
decreased with the size of fluctuation and was in the order of
magnitude of 1 time unit (up to fluctuations of 4 ) (Fig.
\ref{flukt1}b).  The opinion index decayed with approximately the
same characteristic time. On the other hand the fraction of
chartists (which is closely related to volatility) decayed much
slower with characteristic times in the order of magnitude of 100
time units.

\begin{figure}[ht]
\centerline{
\includegraphics[width=14.0truecm]{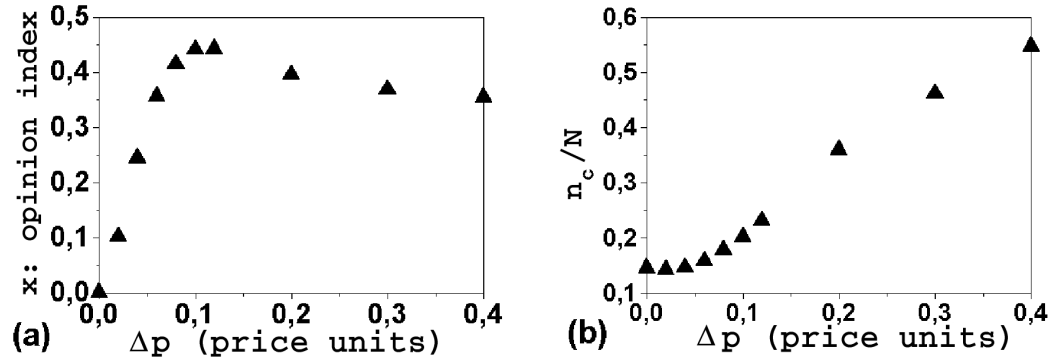}}
\caption{{\protect\footnotesize (a) The average initial opinion
index and  (b) the average initial fraction of chartists as a
function of size of the fluctuation it caused (data points
computed as an average of 7625 to 120529 fluctuations)}}
\label{flukt2}
\end{figure}

An interesting question is what causes fluctuations.  In our
simulations we tried to find an answer by computing the average
opinion index and the average fraction of chartists that caused a
fluctuation as the function of the size of the fluctuation. We
found that small fluctuations are caused merely by the lack of
balance inside the chartist community (Fig. \ref{flukt2}a) while
larger fluctuations are likely to occur only if the fraction of
chartists to all agents rises as well (Fig. \ref{flukt2}b)
resulting in a higher market volatility \cite{lux2}). In other
words the fluctuation in the agents' behaviors results in high
volatility, i.e. in a nervous market which occasionally leads to
major deviations from the fundamental price.  However, in such
cases the attractive force of the fundamental price does not show
up abruptly, since the re-stabilization of the equilibrium price
has to be accompanied by a gradual restoration of the balance
between chartists and fundamentalists.

\section{Response to external effects}

It is well known that financial markets are exposed to many
external effects from the outside world. This means that the
changes in prices are only partly due to the inherent market
mechanisms (fluctuations around "equilibrium"), changes can also
be caused by news from the outside world (e.g. financial reports;
bankruptcy; death of important personalities; outbreak of war;
terror attacks, etc.). The analysis of real market data from the
point of view of external news is a highly non-trivial task. On
the one hand, it is difficult to set an independent level of
"importance" of news in our age of information explosion;
furthermore, the effects of different news may overlap. On the
other hand, the reaction of the market to the news is also hard to
tell. The situation is much simpler in an artificial market like
the Lux-Marchesi model where we can immediately change the
fundamental value of a company or asset (which would be in reality
a consequence of the external event).

\begin{figure}[ht]
\centerline{
\includegraphics[width=14.0truecm]{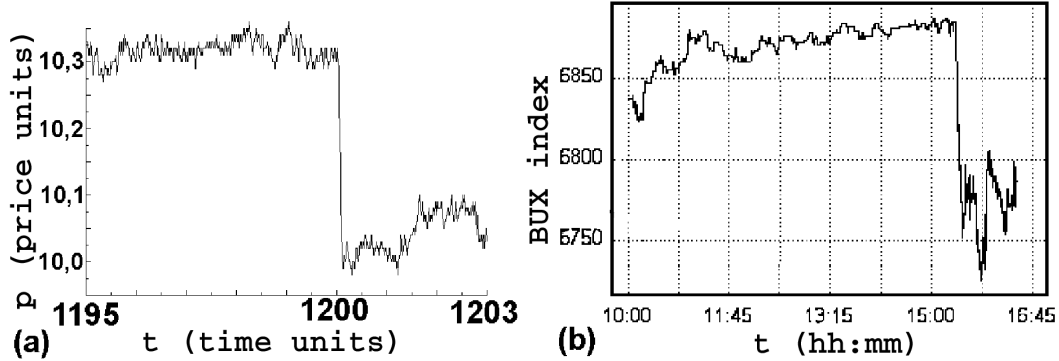}}
\caption{{\protect\footnotesize (a) Price evolution in case of a
fundamental price drop on simulated data (b) and the same on real
market data: Budapest Stock Exchange Index (BUX) drop after the
crash of flight AA587 (early rumors about terror attack) in New
York at 15:17 on 12th November 2001. }} \label{eses1}
\end{figure}

\begin{figure}[ht]
\centerline{
\includegraphics[width=14.0truecm]{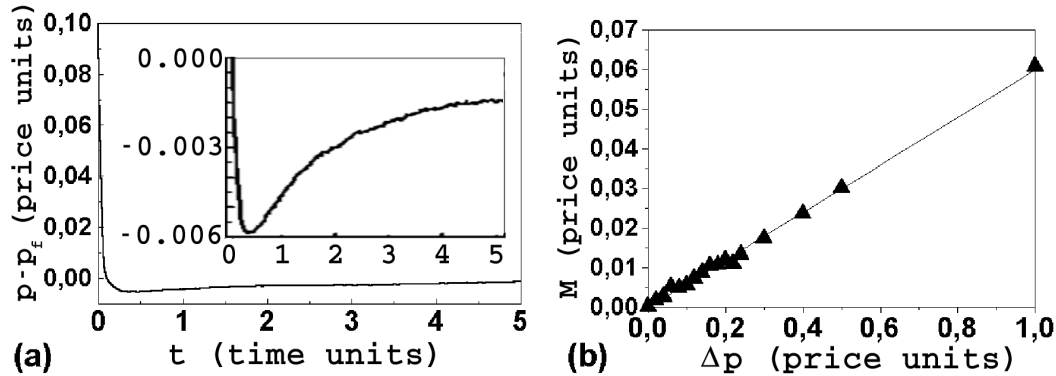}}
\caption{{\protect\footnotesize (a) An average of 50000 price
drops following an abrupt fundamental price change from 10.1\$ to
10.0\$; inset shows the overshoot on a different scale (b) average
size of overshoot as a function of the fundamental price drop
(each data point calculated as an average of 50000 price drops).}}
\label{eses2}
\end{figure}

In our simulation  we changed the fundamental price from
$p_f=10.0+\Delta p$ to $p_f=10.0$ and examined the average of many
runs as a function of time that has passed since the event.  In
the computer program we solved the averaging by raising the
fundamental price by $p$ for 100 time steps and then decreasing it
to the original value, after another 100 time units we raised it
again etc. This means we used a rectangular function
($\Theta$-like function) to perturb the system and recorded the
average response (Fig. \ref{eses2}a). What we saw in case of $p>0$
is an abrupt drop in market price (the speed of which was only
limited by the minimum time step of the model) followed by an
overshoot. Economists and traders have long known that a
correction exists after a very fast price change(Fig.
\ref{eses1}). A sad example for this was the reaction of the
European stock markets to the terror attack on New York on the
11th September. The prices dropped  fast that day (generally
losses over 10 \% were recorded) but the next day there was
already an upward moving trend (a correction after the
overreaction of the events).

We used two different parameter sets (II and IV) to check whether
the occurrence overshoot or its shape depend on the parameters  and  saw
that it is a rather robust effect.
A detailed survey of the phenomenon was undertaken
using parameter set IV.  We saw that the speed of the drop in
price was only limited by the model (this means a maximum of 0.05
price drop in 0.01 time steps).

Another interesting result is that (in case of a price drop) the
location of the price minimum in time is independent of the price
drop $\Delta p$ for a wide range of $\Delta p$ (for parameter set
IV the price minimum is located at approximately $t=0.42\pm 0.02$
time units up to price drops of 10\%). This means that on average
one can predict when the minimum of the price occurs (if one knows
the parameters of the market) irrespective of how big the
fundamental price change $\Delta p$ is.

Let us define the magnitude $M$ of the overshoot as the difference of
the price minimum and the new equilibrium price (which equals the
fundamental price after the event). $M$ shows linear dependence on the
fundamental price change $\Delta p$ in a wide range of $\Delta p$-s
(up to a 10\% abrupt fundamental price change, which is already huge
on market scales) (Fig.  \ref{eses2}b).

When examining the cause of overshoots within the model, the
explanation is at hand: during the sudden drop of the price the
proportion of pessimists rises sharply within the chartists. On
the other hand, the fraction of chartists among all dealers does
not change. This means that the overshoot is caused only by the
movements inside the chartist "community". When the price first
reaches the new equilibrium state, the fraction of pessimists is
still very high which implies the further drop of price, resulting
in the overshoot. Furthermore the opinion index decays to zero
(after a sharp drop immediately after the event) exponentially.

\section{Discussion}

The main result of the presented simulations is that in the
Lux-Marchesi model of financial markets fluctuations and response
to external perturbations (events) decay in a significantly
different manner. Spontaneous fluctuations decay with a relatively
long characteristic time while the response to external events is
practically immediate and followed by an overshoot.  The absence
of the validity of the Onsager hypothesis for this model (and for
financial markets in general) is not at all surprising since the
continuous competition for profit (better payoffs) works so as to
undermine the detailed balance and time reversal symmetry. The
violation of the time reversal symmetry has further consequences as well, 
like the assymetry of the time dependant cross correlations between different
stocks \cite{KKK}.

The investigation of the opinion index and the fraction of
chartists in case of fluctuations and responses showed that 
the decay mechanisms are indeed different:
large spontaneous deviations from the equilibrium price occur in
highly volatile markets which are accompanied by an increase of
the ratio of chartists (pessimists and optimists) as compared to
fundamentalists. The reaction of the market to an external change
of the fundamental price is mainly governed by the shift in the
ratio of the optimists and pessimists inside the group of
chartists. The consequence of the latter is a well defined and
rather robust overshoot in the price. The characteristic time is
short and the size of the overshoot is small indicating that the
market tries to adjust to the new situation effectively, however,
at the same time, the phenomenon itself shows the limitations of
this efficiency.

We have demonstrated that it is worth and possible to investigate the
effect of external perturbations and spontaneous fluctuations
separately in a model market. It would be most interesting to try to
identify the origins of deviations from average behavior on real
market data.  Clearly, several difficulties have to be faced when
trying to distinguish between the two mentioned mechanisms: The
motivation of the agents is hidden, changes are not necessarily such
abrupt as in the model, insider information may influence the pattern
\cite{insider}, effects of different news overlap, large fluctuations
cover the overall behavior, etc. Nevertheless, we believe that our study
gives a hint how to approach this problem.

\begin{ack}
Thanks are due to Thomas Lux for correspondence.  Furthermore we
kindly thank L\'aszl\'o Kullmann for his advice and J\'anos
T\"or\"ok for his help concerning the computer simulations and
other computer related problems. Support of OTKA T029985 is also
acknowledged.
\end{ack}

\end{document}